# A Real-time Image Reconstruction System for Particle Treatment Planning Using Proton Computed Tomography (pCT)

Caesar E. Ordoñez[a]*, Nicholas Karonis[b], Kirk Duffin[b], George Coutrakon[c],
Reinhard Schulte[d], Robert Johnson[e], Mark Pankuch[f]

[a]*Center for Research Computing and Data, Northern Illinois University, DeKalb, IL 60115, USA
[b]Department of Computer Science, Northern Illinois University, DeKalb, IL 60115, USA
[c]Department of Physics, Northern Illinois University, DeKalb, IL 60115, USA
[d]Department of Physics, University of California Santa Cruz, Santa Cruz, CA 90565, USA
[e]Division of Radiation Research, Loma Linda University, Loma Linda, CA 92354, USA
[f]Medical Physics, Northernwestern Medicine Chicago Proton Center, Warrenville, IL 60555, USA

**Abstract**

Proton computed tomography (pCT) is a novel medical imaging modality for mapping the distribution of proton relative stopping power (RSP) in medical objects of interest. Compared to conventional X-ray computed tomography, where range uncertainty margins are around 3.5%, pCT has the potential to provide more accurate measurements to within 1%. This improved efficiency will be beneficial to proton-therapy planning and pre-treatment verification. A prototype pCT imaging device has recently been developed capable of rapidly acquiring low-dose proton radiographs of head-sized objects. We have also developed an advanced, fast image reconstruction software based on distributed computing that utilizes parallel processors and graphical processing units. The combination of fast data acquisition and fast image reconstruction will enable the availability of RSP images within minutes for use in clinical settings. The performance of our image reconstruction software has been evaluated using data collected by the prototype pCT scanner from several phantoms.

## 1. Introduction

Interest in particle-based therapy has recently been increasing due to its potential for highly-targeted dose delivery to organs of interest. To achieve this potential, it is necessary to know the relative stopping power (RSP) of these particles in material media. For proton treatment planning, the RSPs are obtained currently from calibration procedures using X-ray computed tomography (XCT) data. The uncertainties in these indirectly-derived RSPs are about 3.5%. Even such small errors can be critical in accurately delivering dose to desired locations.

Systems for directly measuring proton RSP have been proposed and developed (Schulte, 2004; Sipala, 2013). The pCT device of Schulte, *et al.*, suitable for scanning head-sized objects, has been improved recently to achieve fast tracking and detection of protons (Johnson, 2017). The three-dimensional distribution of RSP in these objects are then obtained via suitable image reconstruction of the acquired proton data (Penfold, 2010).

A major consideration in pCT is that the path of a proton in material media is not a straight line. Thus, conventional fast algorithms for image reconstruction in imaging modalities that use gamma-rays cannot be used. Image reconstruction in pCT must then be done on an event-by-event basis. Typically, good-quality pCT images can be derived from a few hundreds of millions of detected protons. On a single processor, this procedure can take a long time, about an hour.

We have recently reported on accelerated pCT image reconstruction via two levels of parallelization (Karonis, 2013). The first level of parallelization is distributing the data among multiple processors. The second level of parallelization occurs in each individual processor: because proton histories are independent of each other, we used graphical processing units (GPU) to process these histories simultaneously. With this approach, we demonstrated the feasibility of obtaining RSP images from two billion protons in under ten minutes.

In this paper, we present performance results of our improved and optimized pCT reconstruction software, applied to actual data from imaging phantoms measured with the prototype pCT scanner.

## 2. Image Reconstruction

The basic setup for a pCT scanner consists of two pairs of trackers and a range detector in a linear configuration. The object to be imaged is positioned between the two tracker pairs. Protons enter the imaging region through the tracker pair farthest (upstream) from the range detector. The upstream tracker pair measures entry positon and direction into the imaging region. Similarly, the other (downstream) tracker pair measures the exit position and direction from the imaging region. The range detector measures the residual energy of the proton, from which, via a calibration procedure, the water-equivalent path length (WEPL) of the proton as it passed through the object in the imaging region can be calculated. From the detected entry and exit positions and angles, methods have been proposed to estimate the most-likely path (MLP) that each proton traversed in the object (Williams, 2004; Schulte, 2008; Erdelyi, 2009).

The goal of pCT is to derive the 3D distribution of RSP in an object. In practice, this distribution is a discrete rectangular grid of $M$ voxels. The reconstruction problem can then be cast as finding an optimal solution to an over-estimated matrix equation $Ax = b$, where $b$ is an $N$-element vector comprising WEPL values for $N$ (about $10^9$) protons, and $x$ is an $M$-element linearized array of the distribution. $A$ is an $M \times N$ matrix, where each column corresponds to the MLP of each proton through the reconstruction volume. The values in each column will be path lengths in voxels touched by the proton, 0 otherwise. In general, the number of voxels touched by each proton is a very small fraction of $M$, on the order of a few hundred out of about one or two million voxels. Thus, $A$ is a large but sparse matrix, and iterative methods are more suitable for solving the matrix equation (Penfold, 2010).

Here we present only a brief description of image reconstruction of proton data. The implementation of two iterative algorithms in our software is discussed in (Karonis 2013, and references therein). These two algorithms are called diagonally-relaxed orthogonal projections, or DROP (Aharoni, 1989); and component-averaged row projections, or CARP (Gordon, 2005). Figs. 1 and 2 show the block diagram for these two algorithms.

*2.1. DROP*

The data are initially distributed among a user-determined number of blocks. Starting with an initial guess for the solution $x_0$, a new solution $x_1$ is derived by using just the WEPL values $b_0$ and the corresponding sub-matrix $A_0$ for the protons in the first block. The solution $x_1$ is then used as the initial solution for the next block, etc. After the last block is processed, a decision is made whether to repeat the process (new iteration) or stop. Currently, this stopping criterion is set by a fixed, user-defined number of iterations, typically 10. If a new iteration is desired, the last solution $x_N$ is used as initial solution $x_0$ for the next iteration.

*2.2. CARP*

The data are also initially distributed among a user-determined number of blocks (strings in CARP terminology). In contrast to DROP, however, we proceed to get a new estimate $x_n$ for each data subset $n$ using the same initial guess $x_0$. A new solution is then computed from the average of the different string solutions. The stopping criterion is applied: if a new iteration is desired, the initial solution for all the subsets will be this average. CARP is highly suitable for parallel implementation, where the same number of processors can be used as there are processors.

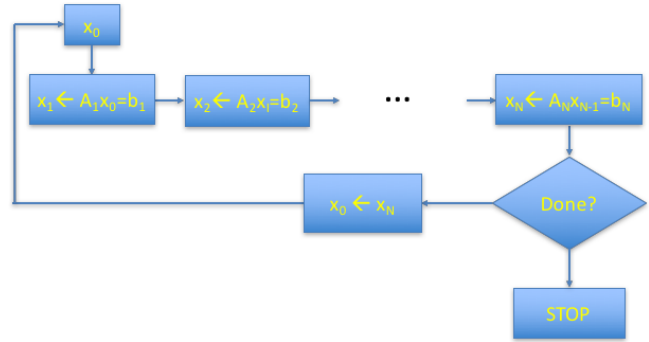

Fig. 1. Schematic diagram of DROP algorithm.

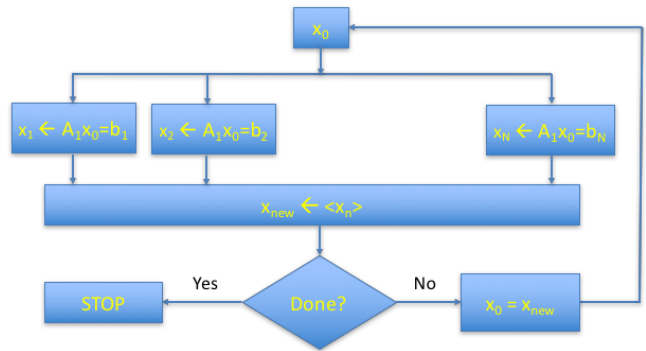

Fig. 2. Schematic diagram of CARP algorithm.

## 3. Materials and Methods

We investigated the performance of our reconstruction software based on (1) execution time and (2) accuracy of the reconstructed images. All reconstruction work was done on the Gaea computer cluster of the Center for Research Computing and Data, Northern Illinois University. The cluster comprises 60 compute nodes and 200 terabytes of storage. Each compute node has two 6-core CPUs and 72 GB RAM; and two Tesla graphical processing units (GPU), each with 6 GB RAM.

*3.1. Execution Time Performance*

Here we used the same approach as our first evaluation of our software (Karonis, 2013). Proton data for 131 million protons obtained by an earlier prototype of the pCT head scanner were used one or more times to simulate data sets with different numbers of proton histories. For this study, total execution time using CARP was investigated as a function of number of protons (problem size) and number of processors (problem scale).

In the discussion below, one processor will denote one CPU core plus one GPU.

*3.2. Quantitative Accuracy*

The prototype pCT scanner was used to image several head-sized phantoms at Northwestern Medicine Chicago Proton Center, Warrenville, IL, using 200-MeV protons. Data from a six-minute scan of the Catphan 404 phantom, continuously rotated at one revolution per minute, were reconstructed using CARP and DROP. This phantom is a plastic cylinder with several cylindrical inserts with known "true" RSPs. The location and names of the visible inserts are shown in Fig. 3. Measured RSPs were extracted from the reconstructed images using region-of-interest (ROI) analysis. The ROI for each material is also a cylinder, with radius and length chosen to minimize partial-volume effects near the walls and ends of the cylindrical material inserts.

## 4. Results

*4.1. Execution Time Performance*

Table 1 summarizes the execution-time performance of our reconstruction software. The first column lists the number of processors used. Columns two to seven show the corresponding execution time (in minutes:seconds) for different numbers of input proton histories. These numbers were chosen to be 1, 2, 4, 8, 12, and 16 times the number of protons in the original data set (131 million).

The baseline times are represented by the row with one processor. Memory limitations prevented successful reconstruction of one billion and higher number of protons using a single processor. Table 1 shows that for a fixed number of processors, execution time increases linearly with the number of protons. We also observe the efficient weak scaling reported in (Karonis 2013). Examination of columns two to six, where the number of histories are multiples (1, 2, 4, 8, 12) of the base set, reveals nearly identical times, shown in boldface, for number of processes in the same multiplicity order.

It is noteworthy that the execution times in the present study are almost exactly a factor of two better than were first reported in (Karonis 2013). The significant improvement results from better optimization of our reconstruction code, as well as implementation of most of the calculations in GPU.

Table 1. Execution-time performance (CARP, time in minutes:seconds).

| Number of Processors | Number of Proton Histories (in billions) | | | | | |
|---|---|---|---|---|---|---|
| | 0.131 | 0.253 | 0.527 | 1.054 | 1.580 | 2.107 |
| 1 | 22:58 | 45:50 | 91:43 | X | X | X |
| 10 | **2:26** | 4:49 | 9:30 | 18:56 | 28:18 | 37:33 |
| 20 | 1:17 | **2:24** | 4:47 | 9:30 | 14:10 | 18:55 |
| 40 | 0:40 | 1:16 | **2:26** | 4:49 | 7:10 | 9:36 |
| 60 | 0:30 | 0:53 | 1:41 | 3:16 | 4:55 | 6:28 |
| 80 | 0:25 | 0:42 | 1:19 | **2:31** | 4:02 | 5:00 |
| 100 | 0:21 | 0:36 | 1:07 | 2:06 | 3:06 | 4:07 |
| 120 | 0:19 | 0:33 | 0:59 | 1:49 | **2:41** | 3:32 |

*4.2. Quantitative Accuracy*

The prototype pCT scanner was used to image several head-sized phantoms at Northwestern Medicine Chicago Proton Center, Warrenville, IL, using 200-MeV protons. Data from a six-minute scan of the Catphan 404 phantom were reconstructed using 60 processors (60 CPU cores + 60 GPUs) on NIU Gaea, with both CARP and DROP. This phantom is a plastic cylinder with several cylindrical inserts with known "true" RSPs. The "true" RSPs were obtained from Peakfinder measurements, except that for air, which was calculated (courtesy of V. Giacometti, University of Wollongong).

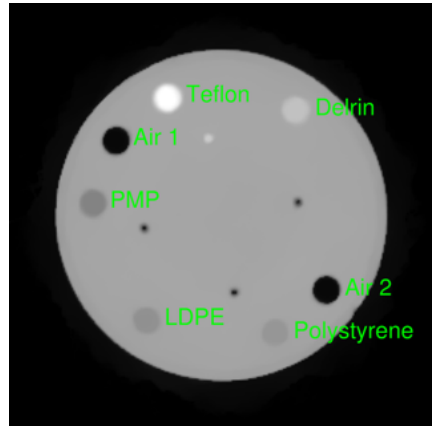

Fig. 3. Middle transverse slice of Catphan 404 phantom showing inserts of known materials.

The means and standard deviations of the inserts' RSPs from region-of-interest (ROI) analysis on the reconstructed images are shown in Table 2. The results for DROP are shown in columns three and four, for CARP in columns six and seven. Also shown in columns five and eight, respectively for DROP and CARP, are the percent deviations of the pCT-measured RSPs from the "true" values.

Table 2 shows that CARP performs better than DROP in the measurement of mean RSPs, achieving better than 1% accuracy for all the material inserts except for air. An advantage for DROP, however, is the overall less noise, as shown by the lower standard deviations compared to those from CARP. This arises from the fact that both algorithms incorporate total variation superiorization (TVS), which "jiggles" the solution to the imaging equation at some point in the reconstruction stream in order to find an optimal, feasible solution. A consequence of TVS is smoothing. In CARP, TVS is applied just once to the initial solution at the start of each iteration. In DROP, however, TVS is applied to the initial solution that enters each block. Therefore, there are as many TVS operations as there are blocks. This results in smoother DROP images, as shown in Fig. 4.

Table 2. Accuracy of reconstructed RSPs.

| Material | True RSP | DROP | | | CARP | | |
|---|---|---|---|---|---|---|---|
| | | Mean | Std Dev | %Error | Mean | Std Dev | %Error |
| Teflon | 1.790 | 1.774 | 0.003 | -0.89 | 1.782 | 0.019 | -0.44 |
| PMP | 0.883 | 0.897 | 0.005 | 1.59 | 0.882 | 0.016 | -0.11 |
| LDPE | 0.979 | 0.986 | 0.003 | 0.72 | 0.971 | 0.018 | -0.82 |
| Polystyrene | 1.024 | 1.031 | 0.006 | 0.76 | 1.017 | 0.018 | -0.68 |
| Delrin | 1.359 | 1.349 | 0.003 | -0.74 | 1.353 | 0.015 | -0.44 |
| Air 1 | 0.00113 | 0.060 | 0.003 | - | 0.029 | 0.010 | - |
| Air 2 | 0.00113 | 0.058 | 0.003 | - | 0.028 | 0.012 | - |

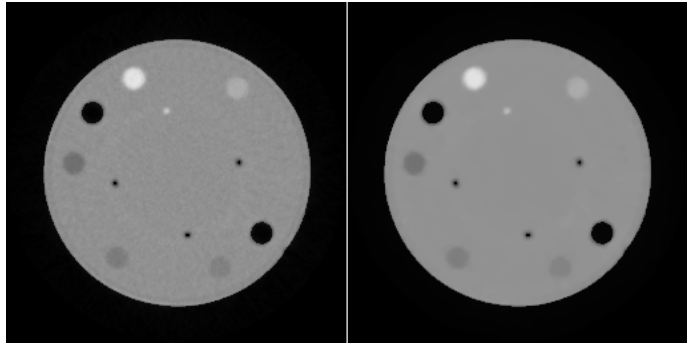

Fig. 4. Comparison of CARP (left) and DROP (right) reconstructed images.

*4.3. Image Quality*

Fig. 5 shows the three middle cardinal slices of pCT reconstruction of a pediatric head phantom, a customized Model 715HN from Computerized Imaging Reference Systems (CIRS). Because the height of the phantom is larger than the vertical extent of the pCT scanner, the superior and inferior halves were scanned separately. The images in Fig. 5 are spliced from separate reconstructions of the superior and inferior halves. The vertical offsets were chosen to have sufficient overlap of transverse slices. The images clearly show the internal structures of the phantom. The square in the coronal slice, the rectangles in the sagittal and transverse slices, show the presence of film in the removal insert. The outline of the cavity for the insert is also visible, as well as the gap between the two halves of the phantom.

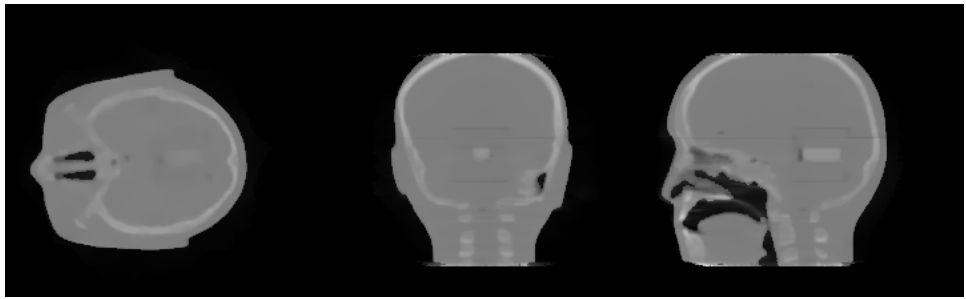

Fig. 5. Transverse (left), coronal (middle), and sagittal (right) slices of pCT reconstruction of pediatric head phantom (CIRS customized Model 715HN).

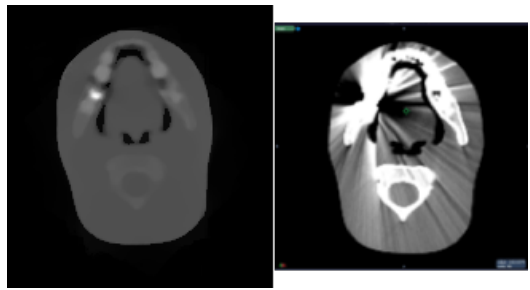

Fig. 6. Transverse slices from pCT (left) and XCT (right) through gold tooth in pediatric head phantom.

Fig. 6 shows another big advantage of pCT over XCT. The pediatric head phantom has a couple of gold molars. The figure shows approximately the same transverse slices in pCT and XCT that cut through the gold tooth. The XCT image shows severe streaking artifacts due to the effects of the high-Z material on gamma-rays. In contrast, no such artifacts are observed in the pCT images.

## 5. Conclusions

We have demonstrated that fast, accurate, and good-quality images can be achieved with proton computed tomography. Reconstruction times are of the order of a few minutes, comparable to the time needed to acquire data with sufficient statistics to yield artifact-free images. We are investigating streamlining the data acquisition, data preprocessing, and pCT reconstruction so that all could be done in the clinic in real time.

Both DROP and CARP algorithms yield satisfactory reconstructions of RSPs in head-sized objects in this study. RSPs obtained with CARP tend to be more quantitatively accurate.